\documentclass[twocolumn,showpacs, superscriptaddress,aps]{revtex4}
\usepackage{graphicx}
\begin{document}
\title{Quantum phase transition and underscreened Kondo effect in electron transport through
parallel double quantum dots}
\author { Guo-Hui Ding}

 \affiliation{Department of Physics, Shanghai Jiao Tong
University, Shanghai 200240, China}
\author {  Fei Ye}
\affiliation{ Center for Advanced Study, Tsinghua University,
Beijing 100084, China}
\author { Bing Dong}
 \affiliation{Department of Physics, Shanghai Jiao Tong
University, Shanghai 200240, China}

\date{\today }

\begin{abstract}
  We investigate electronic transport through parallel double quantum
dot(DQD) system with strong on-site Coulomb interaction and
capacitive interdot coupling. By applying numerical renormalization
group(NRG) method, the ground state of the system and the
transmission probability at zero temperature have been obtained. For
a system of quantum dots with degenerate energy levels and small
interdot tunnel coupling, the spin correlations between the DQDs is
ferromagnetic, and the ground state of the system is a spin $1$
triplet state. The linear conductance will reach the unitary limit
($2e^2/h$) due to the underscreened Kondo effect at low temperature.
As the interdot tunnel coupling increases, there is a quantum phase
transition from ferromagnetic to anti-ferromagnetic spin correlation
in DQDs and the linear conductance is strongly suppressed.
\end{abstract}

\maketitle

\newpage

\section{introduction}
  In recent years considerable research attention has been paid to
electron transport through double quantum
dot(DQD)systems\cite{Wiel}, which are an artificial small quantum
systems can be readily controlled by external gate voltage and also
exhibit a variety of interesting strongly correlated electron
behaviors. Basically, there are two different experimental
realizations of DQD system: DQDs connected in serial\cite{Blick} or
in parallel configurations\cite{Chen}. Electron transport through
both configurations have been studied in experiments,  the molecular
states of the double dots and also the competition between Kondo
effect and the RKKY interaction have been observed\cite{Blick,Chen}.

 The theoretical studies on electron transport through DQDs are largely
devoted to the system in the Kondo regime. For DQDs connected in
serial, the antiferromagnetic correlations between two single-level
coupled QDs is in competition with Kondo correlations between the
QDs and the electrons in the leads, therefore it gives rise to rich
ground state physical properties at zero temperature\cite{Georges,
Aguado, Lopez, Mravlje}. For DQDs with large capacitive coupling,
the simultaneous appearance of the Kondo effect in the spin and
charge sectors results in an SU(4) Fermi liquid ground
state\cite{Borda}. By increasing interdot capacitive coupling, a
quantum phase transition of Kosterlitz-Thouless type to a
non-Fermi-liquid state with anomalous transport properties is
predicted\cite{Galpin}. Martins et al argued that ferromagnetic
state cannot be realized in two single-level QDs connected in
serial, but they predicts that FM state can be developed in two
double-level QDs\cite{Martins}. For the DQD system in  parallel
configuration, the physical properties can be quite different, since
the interference effect will play an important role in its transport
properties. The Fano effect for electron transport through bonding
and antibonding channels in DQDs system has been
studied\cite{Guevara,Ding,Tanaka}.

Due to the strong correlation of electrons in the QDs, it is a
non-trivial problem to treat those systems theoretically. It is well
known that Wilson's numerical renormalization group\cite{Wilson,
Krishna-murthy, Costic, Bulla} method is a nonperturbative approach
to quantum impurity problem, which can take into account the on-site
Coulomb repulsion and the spin exchange interaction between the
electrons in  DQDs exactly, in contrast to the slave boson mean
field theory or the equation of motion method within Hartree-Fock
approximation. The NRG method have already been applied to
investigate a lot of problems in the electron transport through QD
systems. For instance, DQDs connected in serial\cite{Izumida}, the
quantum phase transition in multilevel QD\cite{Hofstetter}, Kondo
effect in coupled DQDs with RKKY interaction in external magnetic
field\cite{Chung}, the side coupled DQD system\cite{Cornaglia,Zitko}
and quantum phase transitions in parallel quantum QDs \cite{Zitko1}
etc. However, in our opinion the consequences of the interplay of
Fano resonance and the Kondo effect on electron conductance through
DQDs in parallel still haven't been well elucidated. In this paper
we will investigate the electron transport properties for the DQDs
in parallel configuration by using the NRG method. We will show that
for DQDs without interdot tunneling, the underscreen Kondo effect
plays an essential role in the conductance. The linear conductance,
spin correlation, and local density of state in this system are
obtained.

\section{The model Hamiltonian and the NRG approach}
  Electron transport through parallel-coupled DQDs with interdot
tunneling, on-site Coulomb interaction and capacitive interdot
coupling can be described by the following Anderson impurity model:
\begin{eqnarray}
H&=&\sum_{k\eta\sigma} \epsilon_{k\eta} c^\dagger_{k\eta\sigma}
c_{k\eta\sigma}+\sum_{i\sigma}\epsilon_i
d^\dagger_{i\sigma}d_{i\sigma}+\sum_{i}U
n_{i\uparrow}n_{i\downarrow}+Vn_{1}n_{2} \nonumber\\
&+&t_c\sum_\sigma(d^\dagger_{1\sigma}d_{2\sigma}+d^\dagger_{2\sigma}d_{1\sigma})
+\sum_{k\eta\sigma i} (v_{\eta
i}d^\dagger_{i\sigma}c_{k\eta\sigma}+H.c.)\;,
\end{eqnarray}
where $c_{k\eta\sigma}(c^\dagger_{k\eta\sigma})$ denote
annihilation(creation) operators for electrons in the leads($\eta=L,
R$),  and $d_{i\sigma}(d^\dagger_{i\sigma})$ those of the single
level state in the i-th dot( $i=1,2$). $n_{i\sigma}$ denotes the
electron number operator with spin index  $\sigma$ in the i-th dot,
and $n_i=\sum_\sigma n_{i\sigma}$.   $U$ is the intra-dot Coulomb
interaction between electrons, $V$ is the interdot capacitive
coupling.  $t_c$ is the interdot tunnel coupling, and $v_{\eta i}$
is the tunnel matrix element between lead $\eta$ and dot $i$.  It
should be noted that an interdot magnetic exchange term $J$ is not
explicitly included in this Hamiltonian since it is not an
independent parameter but a function of the interdot
tunneling($J\sim t_c^2/U$). We consider the symmetric coupling case
with $\Gamma_i^L=\Gamma_i^R=\Gamma_i$ , where
$\Gamma_i^\eta=2\pi\sum_k |v_{\eta i}|^2
\delta(\omega-\epsilon_{k\eta\sigma})$ is the hybridization strength
between the $i$-th dot and the lead $\eta$.

In order to access the low-energy physics of this DQD system, we
adopt the Wilson's NRG approach.  By symmetric combination of the
lead orbitals, the Hamiltonian in eq.(1) can be mapped to
single-channel two-impurity Anderson model. Because the
anti-symmetric combination of lead orbitals are totally decoupled
with the QDs, they can be neglected in the Hamiltonian. Following
the standard NRG method, one defines a series of rescaled
Hamiltonian $H_N$ as following
\begin{eqnarray}
H_N&=&\Lambda^{(N-1)/2}[\sum_{\sigma,n=0}^{N-1}\Lambda^{-n/2}\xi_n(f^\dagger_{n\sigma}f_{n+1\sigma}
+f^\dagger_{n+1\sigma}f_{n\sigma})
\nonumber\\
&+&\sum_{i\sigma}(\tilde\epsilon_i+{1\over 2}\tilde
U)d^\dagger_{i\sigma}d_{i\sigma}+{1\over 2}{\tilde U}\sum_i
(n_{i}-1)^2
 +{\tilde V}n_{1}n_{2}
 \nonumber\\
&+&\tilde t_c
\sum_\sigma(d^\dagger_{1\sigma}d_{2\sigma}+d^\dagger_{2\sigma}d_{1\sigma})+
\sum_{i\sigma}{\tilde\Gamma_i}^{1/2}(f^\dagger_{0\sigma}d_{i\sigma}+d^\dagger_{i\sigma}f_{0\sigma})]\;,
\nonumber\\
\end{eqnarray}
where the discretization parameter $\Lambda=1.5$, and $\xi_n\approx
1$\cite{Bulla}. The other parameters
$\tilde\epsilon_i={2\over{1+\Lambda^{-1}}}{\epsilon_i\over D}$,
$\tilde U={2\over{1+\Lambda^{-1}}}{U\over D}$,$\tilde
V={2\over{1+\Lambda^{-1}}}{V\over D}$, $\tilde
t_c={2\over{1+\Lambda^{-1}}}{t_c\over D}$ and
$\tilde\Gamma_i=({2\over{1+\Lambda^{-1}}})^2{\Gamma_i\over{\pi D}}$,
with $D$ being the bandwidth of electrons in the leads.  The above
one dimensional lattice model is iteratively diagonalized by using
the recursion relation
\begin{equation}
H_{N+1}=\Lambda^{1/2}
H_N+\xi_n\sum_\sigma(f^\dagger_{N\sigma}f_{N+1\sigma}
+f^\dagger_{N+1\sigma}f_{N\sigma})\;.\\
\end{equation}
The basis set in each iteration step is truncated by retaining
only those states with low-lying energies. In our numerical
calculation, we keep totally $600$ low-lying energy states in each
step without counting the $S_z$ degeneracy.

 The current formula through the DQDs is given by the generalized
Landauer formula\cite{Meir}
\begin{equation}
I={e\over h}\sum_{\sigma}\int d\omega
[n_L(\omega)-n_R(\omega)]T(\omega)\;,\\
\end{equation}
where the transmission probability $T(\omega)=-Tr[ {\hat\Gamma}
Im[ \hat G^r(\omega)]]$, with
${\hat\Gamma}={\hat\Gamma^L}={\hat\Gamma}^R=
\left ( \begin{array}{cc} {\Gamma}_1 & \sqrt{{\Gamma}_1{\Gamma}_2  } \\
\sqrt{{\Gamma}_1{\Gamma}_2  } & {\Gamma}_2
\end{array} \right )$. The
retarded/advanced  Green's functions(GF) $\hat G^{r/a}(\omega)$ have
$2\times 2$ matrix structures, which account for the double dot
structure of the system. The matrix elements of the retarded GF are
defined in time space as
$G_{ij}^r(t-t')=-i\theta(t-t')<\{d_{i\sigma}(t),
d^+_{j\sigma}(t')\}>$. Therefore, the transmission probability
$T(\omega)$ can be obtained by calculating the imaginary parts of
the GF of DQDs or the spectral density $\rho_{ij}(\omega)=-{1\over
\pi}Im G^r_{ij}(\omega)$. Then, the linear conductance at the
absolute zero temperature can be given by taking the zero frequency
limit of the transmission probability
$G={dI\over{dV}}|_{V=0}={2e^2\over h}T(\omega=0)$. One advantage of
the NRG  is accurate determination of the low-energy spectral
density of the quantum impurity models. By a standard procedure in
NRG \cite{Bulla},  the spectral density at zero temperature can be
calculated according to the following formula
\begin{eqnarray}
\rho_{ij}(\omega)&=&{1\over Z(0)}\sum_\lambda
M^i_{0,\lambda}(M^{j}_{0,\lambda})^*\delta (\omega-(E_\lambda-E_0))
\nonumber\\
&+&{1\over Z(0)}\sum_\lambda M^i_{\lambda,0}(M^{j}_{\lambda,
0})^*\delta (\omega+(E_\lambda-E_0))
\end{eqnarray}
where the matrix element $M^i_{\lambda,0}=<\lambda|d_{i\sigma}|0>$,
 with $|0>$ and $|\lambda>$ being the ground state and exited eigenstate of
the impurity model Hamiltonian, respectively.

\begin{figure}[htp]
\includegraphics[width=0.9\columnwidth, height=3in]{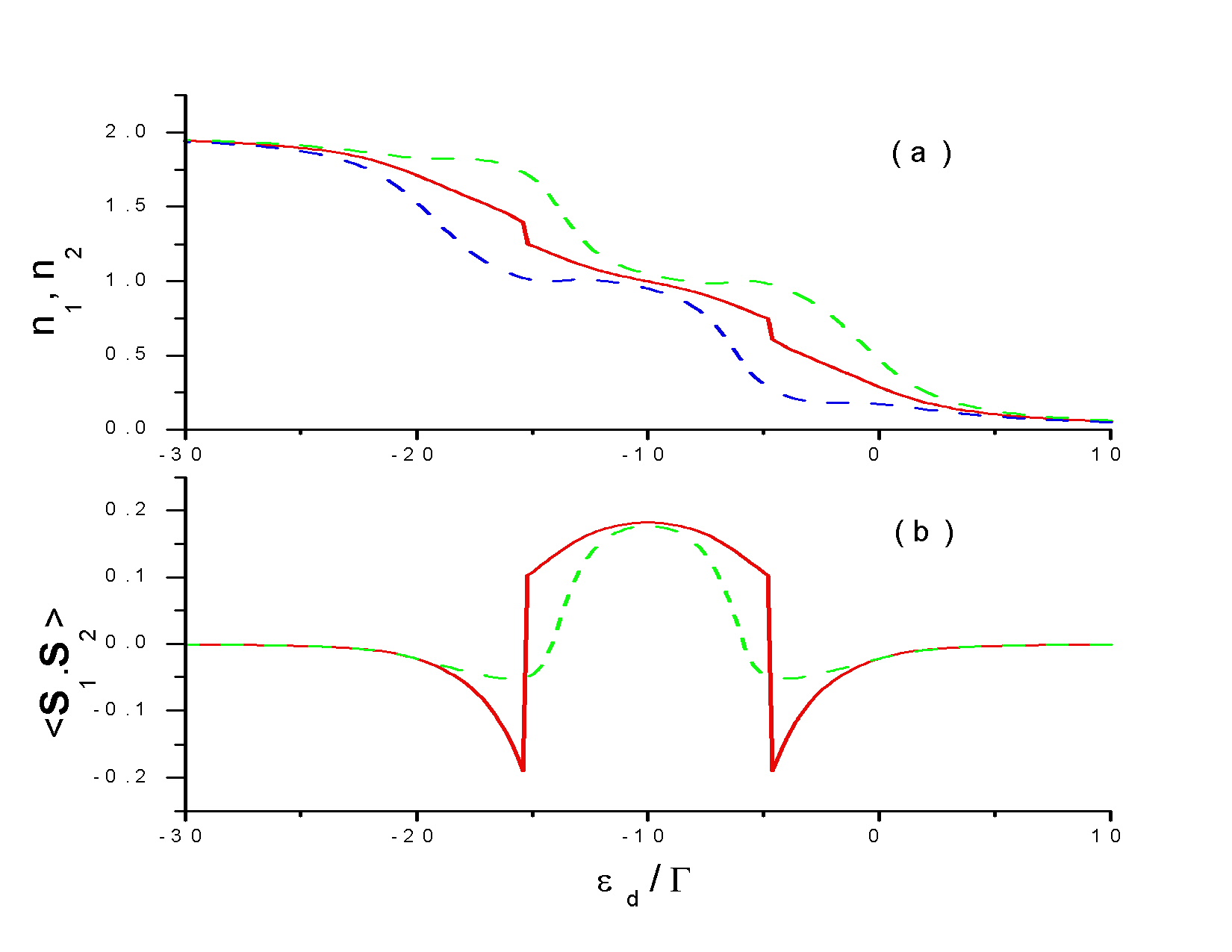}
\caption{ (a) The electron occupation number $<n_{i}>$ in each
quantum dot as a function of the gate voltage $\epsilon_d$.
$\Delta\epsilon_d/\Gamma=0 $(solid line); $2.0$ (dashed line). The
other used parameters are $D=1.0$, $t_c=0$, $\Gamma=0.01$,
$U/\Gamma=10$, and $V=U/2$; (b) The interdot spin correlation $<{\bf
S}_1\cdot {\bf S}_2>$ as a function of the gate voltage
$\epsilon_d$.  }
\end{figure}

\section{results and discussions}
 In the following, we will present the results of our NRG
 calculation. For the sake of simplicity, we only consider the
 symmetric coupling case with the hybridization strength
 $\Gamma_1=\Gamma_2\equiv\Gamma$.   We take the bandwidth $D=1$  as
the energy unit, and the other parameters $\Gamma=0.01$, $U=10
\Gamma$, $V=U/2$. One can define the averaged energy level of QDs as
$\epsilon_d=(\epsilon_1+\epsilon_2)/2$, and the energy level
difference $\Delta \epsilon_d=\epsilon_2-\epsilon_1$. Both of them
can be tuned experimentally by external gate voltages.

At first, we consider DQDs without interdot tunneling ($t_c=0$). In
Fig.1(a) the occupation number of electrons $<n_{i}>$ in each QD is
plotted as a function of the average energy level $\epsilon_d$. The
electron occupation number increases consecutively by tuning the QD
level below the Fermi energy.  For this DQDs with interdot
capacitive interaction, one can easily discern the different regions
of occupation states: from empty occupation to the state with total
four electrons in DQDs . In the case of two identical QDs (
$\Delta\epsilon_d=0$), abrupt jumps of the occupation number are
observed at some particular gate voltage. One can see that the
position of jumps can be identified as the region where the DQDS
have odd number of electrons. For DQDs with different energy
levels($\Delta\epsilon_d\neq  0$), the QD with low energy level is
occupied first, and because of interdot capacitive interaction, it
will greatly suppress the occupation of electron in another QD as
compared with the two identical QDs case. The interdot
spin-correlation $<{\bf S}_1\cdot {\bf S}_2> $ as a function of
energy level $\epsilon_d$ is shown in Fig.1(b), where the spin
operators in the i-th QD are defined by ${\bf
S}_i=1/2\sum_{\sigma\sigma'}d^\dagger_{i\sigma}{\bf\sigma}_{\sigma\sigma'}d_{i\sigma'}$.
It shows that the interdot spin correlation is antiferromagnetic in
the mixed valence regime,  and is ferromagnetic in the doubly
occupied regime, where each QD is occupied by one electron. For DQDs
with energy level difference, the spin correlation in mixed valence
regime is greatly suppressed, but there are still large
ferromagnetic spin correlation in the doubly occupied regime. In the
identical QDs($\Delta\epsilon_d=0$ ) case,  the abrupt jumps in
occupancy and the spin correlation turn from FM to AFM have also
been found in Ref.\cite{Zitko1} for $N$-QD system($N\geq 2$) without
interdot capacitive coupling, and this phenomenon is interpreted as
a kind of quantum phase transitions. However, one can see from the
dashed line in Fig.1 that the abrupt jumps both in occupancy and
spin correlation disappear when $\Delta\epsilon_d\neq 0$, hence this
kind of phase transition is unstable with respect to the
perturbation by gate voltage difference in QDs. We attribute this
kind of abrupt jump as a result of  Fano resonance and the crossing
of the antibonding state energy level with the Fermi energy.

\begin{figure}[htp]
\includegraphics[width=0.9\columnwidth, height=3in]{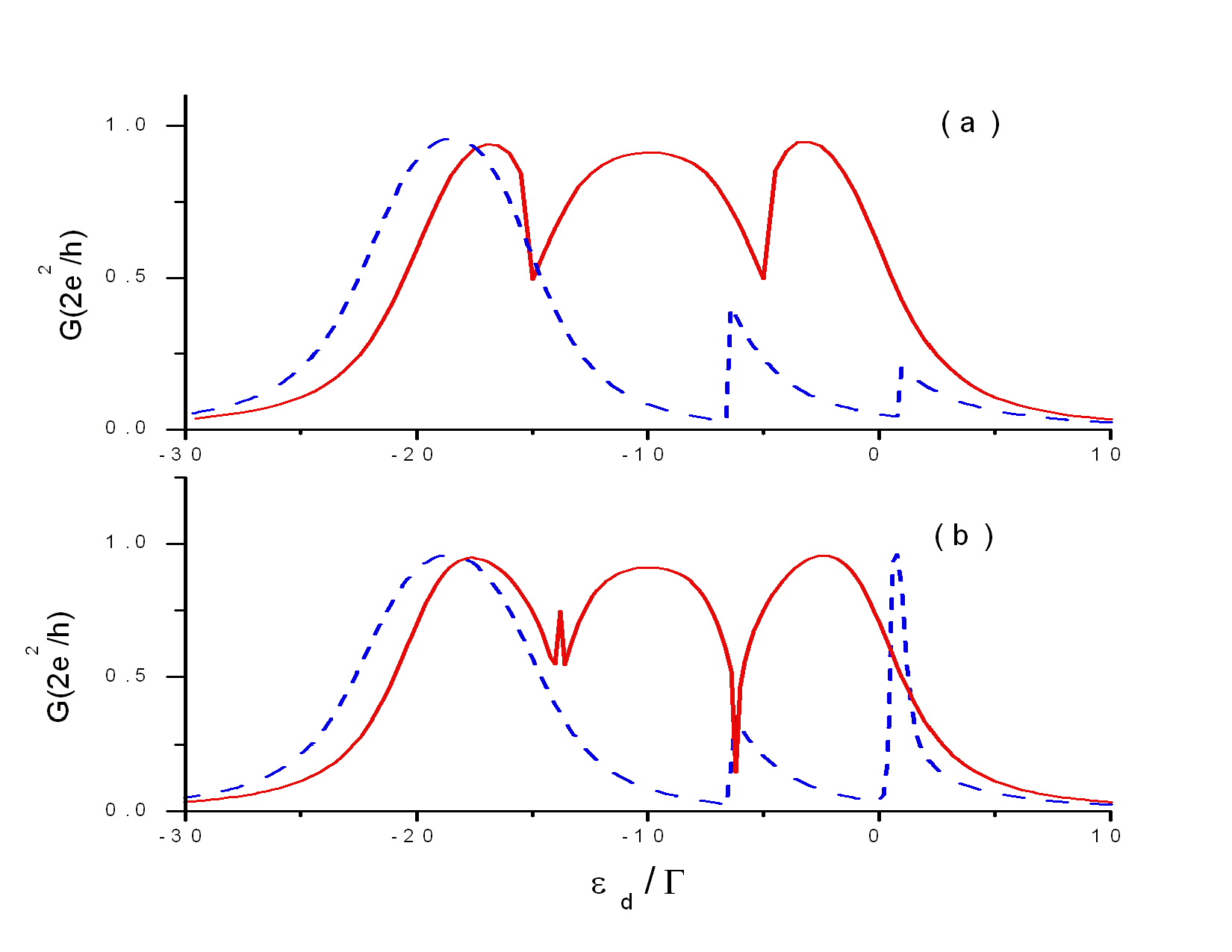}
\caption{The linear conductance $G$ as a function of the dot level
at zero temperature. (a) for the system with two identical quantum
dots($\Delta\epsilon_d=0$); (b) for DQDs with energy level
difference($\Delta\epsilon_d/\Gamma=2.0$).  The interdot tunneling
parameter takes $t_c/\Gamma=0$(solid line) and
$t_c/\Gamma=2.0$(dashed line), respectively.}
\end{figure}

Next, we calculate the electron conductance through DQDs when a
small bias voltage is applied to the leads.  In Fig.2 the linear
conductance $G$ at zero temperature vs. the average QD energy level
$\epsilon_d$ is depicted. As shown in Fig.2(a), the Kondo effects
are manifested by peaks in the curve of the linear conductance,
where the conductances approach the unitary limit ($G=2e^2/h$). In
the regime of odd electron occupation, the DQDs act as a localized
spin (s=1/2), and the Kondo effect is arose from the spin exchange
interaction between the localized electron spin and that of the
electrons in the leads. Whereas, in the doubly occupied regime, the
unitary conductance is due to the underscreened spin 1 Kondo effect,
It will be shown in the following that in this regime the two
electrons confined in the DQDS  form a spin triplet state in the
ground state. In the presence of sufficient interdot tunneling
$t_c$, the Kondo effect in the singly occupied regime and spin 1
Kondo effect is strongly suppressed, but some asymmetrical peaks of
conductance appears in the mixed valence regime, this can be
attributed to the Fano resonance for the electron transport through
the bonding and antibonding channels in this system. It is
interesting to notice that in the triply occupied regime the
conductance still achieves the unitary limit even in the presence of
strong interdot tunnel coupling.

\begin{figure}[htp]
\includegraphics[width=0.9\columnwidth, height=3in]{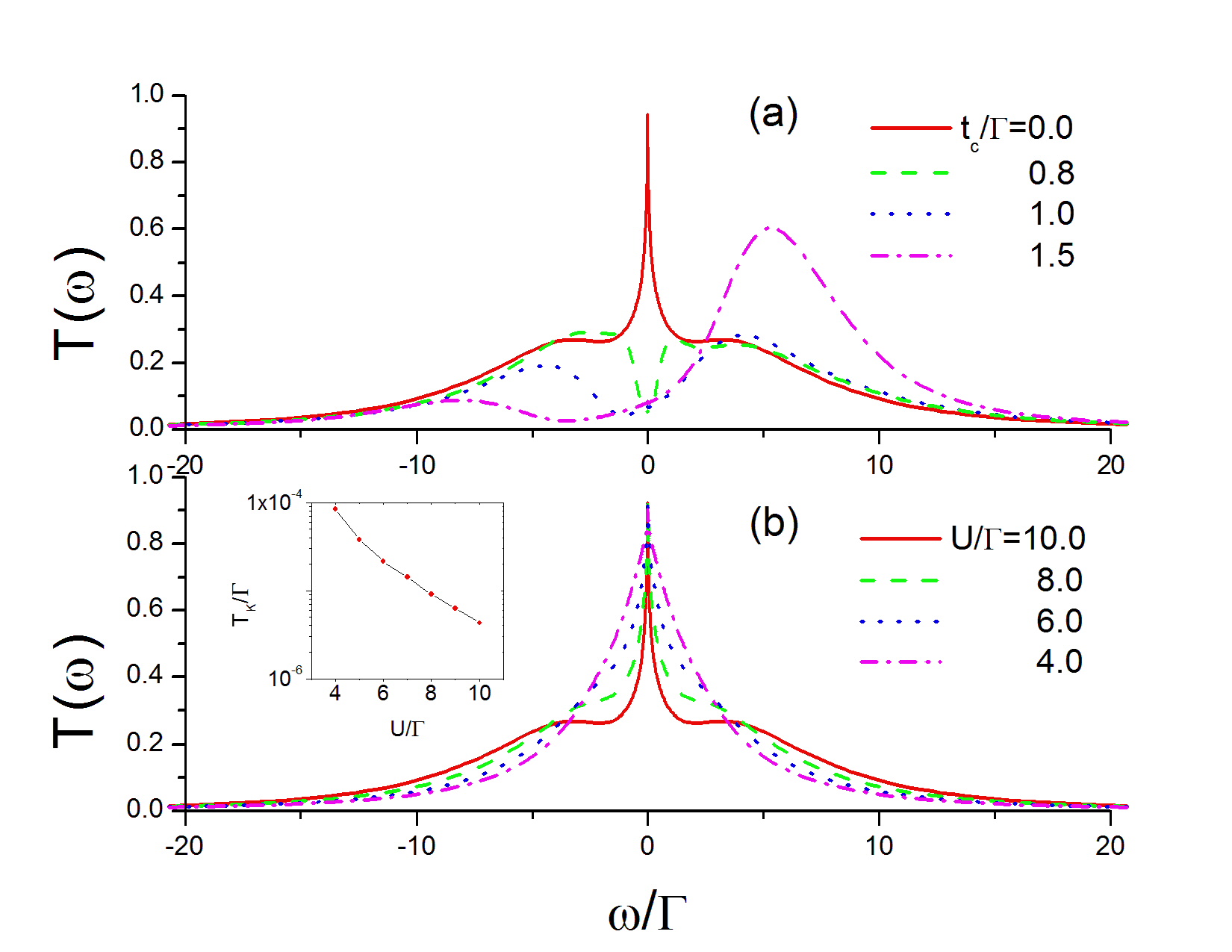}
\caption{(a) The transmission probability $T(\omega)$  for the
system with two identical quantum dots.  Used parameters are
$D=1.0$, $\Gamma=0.01$, $U/\Gamma=10$, $V=U/2$ and
$\epsilon_d/\Gamma=-10.0$. The interdot tunnel coupling $t_c/\Gamma
=0.0, 0.8, 1.0,1.5$, respectively. (b) The transmission probability
$T(\omega)$ at the particle-hole symmetric point for different
values of on-site Coulomb interaction $U$. Inset: the estimated
Kondo temperature $T_K$ vs. the value of $U$. }
\end{figure}

In the following, we will focus our attentions on the properties in
the doubly occupied regime. In order to illustrate the effect of
interdot tunneling, the transmission probability at different
tunneling coupling $t_c$ is shown in Fig.3 (a). Without direct
interdot tunneling ($t_c=0$), one can see that the transmission
probability has the particle-hole symmetry, and the spin exchange
effect between the electrons localized in the quantum dots and that
in the leads gives rise to a sharp peak in the transmission
probability at the Fermi surface, therefore the linear conductance
at zero temperature reaches the unitary limit $G=2e^2/h$, as a
result of the underscreened spin 1 Kondo effect. In the presence of
the interdot coupling $t_c\neq 0$ , the particle-hole symmetry of
the transmission probability is broken. When $t_c$ increases beyond
a quantum critical point, a sharp dip in the transmission
probability is observed. It suggests that the Kondo effect and the
linear conductance in this regime is strongly suppressed. Therefore,
there is a quantum phase transition between underscreened Kondo
phase  and the local spin singlet phase in the ground state of this
system. For two-impurity Anderson model without interdot capacitive
coupling, this quantum phase transition has been predicted by
Nishimoto et al.  by using dynamic density matrix renormalization
group \cite{Nishimoto},  and \v{Z}itko et al. have obtained its
thermodynamic properties, such as the temperature dependence of
magnetic susceptibility and entropy by  NRG method \cite{Zitko2}. It
is noted that a similar quantum phase transition is also observed in
two-level single QD system with intradot spin exchange coupling by
Hund's rule\cite{Hofstetter}. For DQDs with RKKY interaction coupled
to two-channel lead, Chung et al.\cite{Chung} found the quantum
phase transition is from Kondo screened phase to spin singlet phase.
In Fig.3(a), by further increasing the interdot coupling $t_c$, a
broad peak of transmission probability with the line shape of
Breit-Wigner resonance is developed around the energy $\omega\approx
U/2$, we attribute this broad transmission peak to the electron
transport through the bonding channel of electrons in the quantum
dots.

    In Fig.3(b) the transmission probability $T(\omega)$ at
different values of on-site Coulomb interaction  $U$ is depicted. It
shows that the line shape of the $T(\omega)$ changes significantly
by varying the Coulomb interaction strength $U$. The line shape
becomes more cusplike  with decreasing  $U$ , and it reveals that
the physical properties of this underscreened Kondo effect in  DQD
system is quite different from the spin 1/2 Kondo effect. For the
spin 1/2 Kondo effect in single-impurity Anderson model, one can
estimate the Kondo temperature $T_K$ by using the formula
$T_K={\sqrt{U\Gamma}\over
2}exp[{\epsilon_d(\epsilon_d+U)/U\Gamma}]$. For this underscreened
Kondo effect case, we make the following approximation to estimate
the Kondo temperature : at the frequency of $\omega=T_K$ the
transmission probability $T(\omega=T_K)/T(\omega=0)\approx 0.978 $.
For the single-impurity Anderson model, $T_K$ obtained by this
approximation agrees well with the above formula. The inset of Fig.3
(b) shows the estimated  $T_K$ at several values of the Coulomb
interaction strength $U$ for the DQD system. For the system with the
parameters used in our calculation, the Kondo temperature $T_K$ is
on the order of $10^{-5}\Gamma$.

\begin{figure}[htp]
\includegraphics[width=0.9\columnwidth, height=3in]{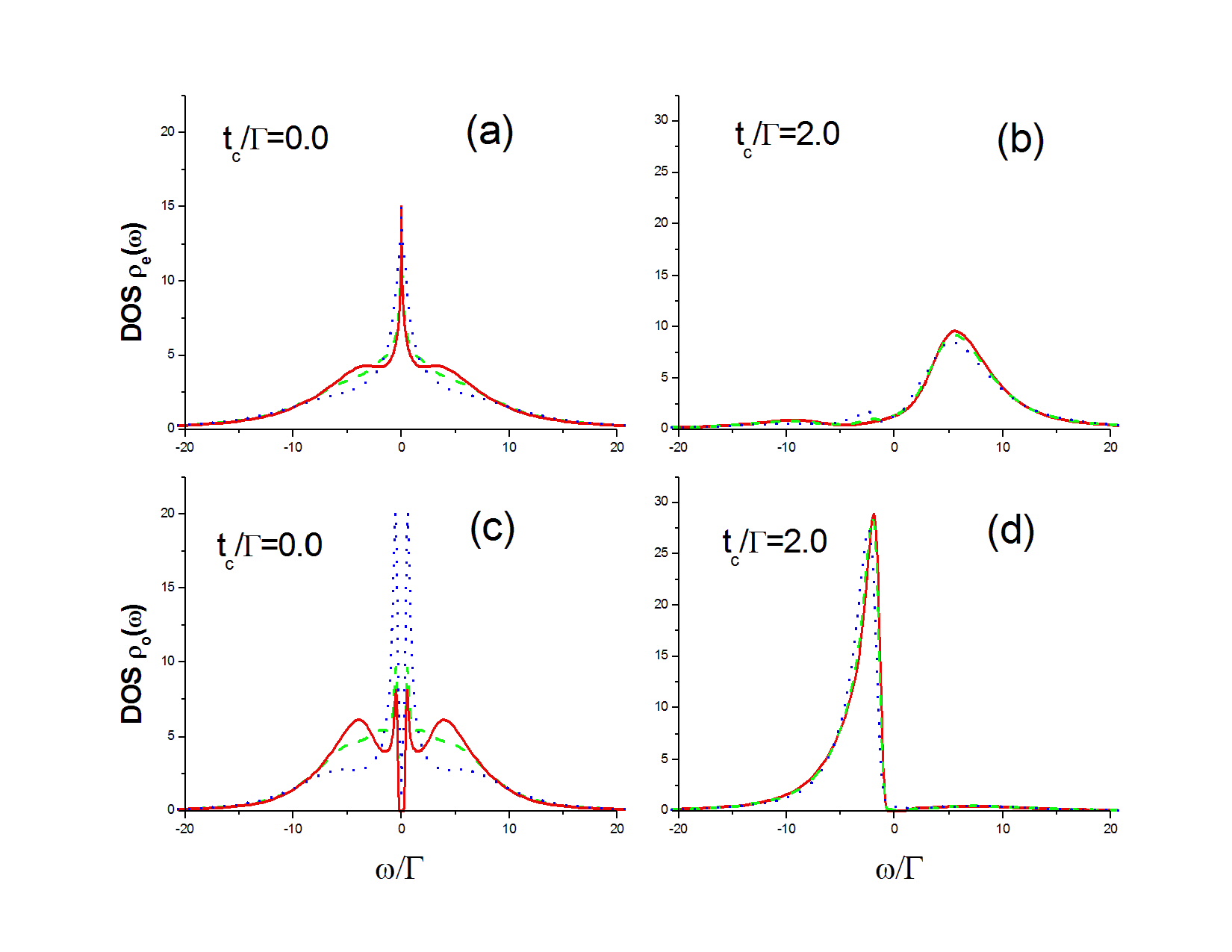}
\caption{ The  density of state of the local bonding and antibonding
states in the quantum dot at different values of energy level
difference: $\Delta \epsilon_d/\Gamma=0.0$(solid line), 2.0(dashed
line), $4.0$(dotted line). (a), (b) corresponds to the bonding state
with $t_c/\Gamma=0,0, 2.0$, respectively. (c),(d) are that of the
antibonding states. The other used parameters are the same as in
Fig.3. }
\end{figure}

In order to get better understanding of the electron state in the
system, we investigate the local density of states(DOS) in the DQDs.
One can define the even orbital (bonding state) operator as
$d_{e,\sigma}=(d_{1\sigma}+d_{2\sigma})/\sqrt{2}$, and the odd
orbital (antibonding state) operator
$d_{e,\sigma}=(d_{1\sigma}-d_{2\sigma})/\sqrt{2}$. The local density
of state for the bonding and antibonding states are depicted in
Fig.4. As shown in Fig.4(a) and (c),   in the absence of interdot
coupling($t_c=0$), the local DOS of even and odd orbital  retain the
particle-hole symmetry of the system. It is noticed that the
transmission probability is proportional to the DOS for the bonding
state, therefore a Kondo peak around the Fermi energy is observed in
its DOS. Some new features are also manifested in DOS for this
system, one can see that the DOS for the antibonding state has two
side peaks nearby the Fermi energy , which can be understood as a
result of the effective spin-exchange interaction between the
electrons in DQDs by tunneling through the leads, and this feature
cannot be found in DQDs in serial configuration\cite{Izumida}. As
the interdot coupling $t_c$ is large than some critical value(see
Fig. 4(b) and (d)), the Kondo effect on the DOS of bonding state is
greatly suppressed, and a broad peak around the energy
$\omega\approx U/2$ are developed.  For the DOS of the antibonding
state,  a sharp peak is developed slight below the Fermi energy,
this is due to the fact that  the antibonding state of electrons in
DQDs  seems like a quasi-localized state. Increasing the interdot
coupling $t_c$ further, the sharp peak is broaden and shifts away
from the Fermi surface to lower energy. For DQDs with different
energy levels,  the characteristic features of the DOS remain
unchanged.

\begin{figure}[htp]
\includegraphics[width=0.9\columnwidth, height=3in]{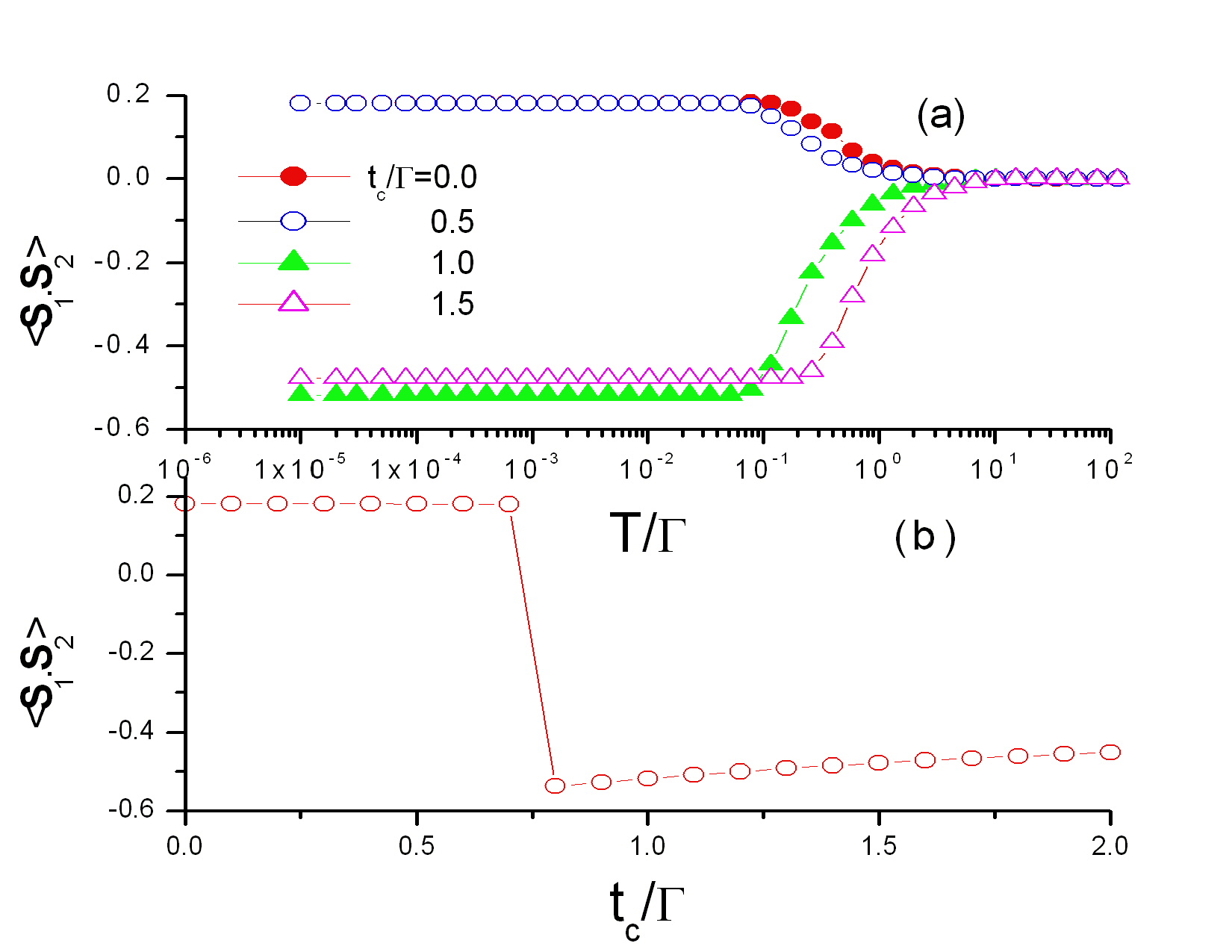}
\caption{(a)The spin correlation $<{\bf S}_1\cdot {\bf S}_2>$ of
double quantum dots as a function of temperature $T$ for several
different values of $t_c$. (b) The spin correlation $<{\bf S}_1\cdot
{\bf S}_2>$ vs. the interdot tunnel coupling $t_c$ at zero
temperature. The other used parameters are the same as in Fig.3. }
\end{figure}

To gain more insight into  the spin entanglement and the effect of
spin exchange interaction for the electrons localized in different
QDs, We have also calculated the interdot spin-correlation $<{\bf
S}_1\cdot {\bf S}_2> $ as a function of temperature for several
values of interdot tunneling $t_c$ as shown in Fig.5(a). When
interdot tunneling $t_c$ is zero or has a small value, one can see
that the spin-correlation converges to a positive value as
temperature decreases. It is easy to notice that the positive value
of $<{\bf S}_1\cdot {\bf S}_2> $ reveals that the spin-correlation
in this case is ferromagnetic type in the ground state. As we know
that when two ideal spin $s=1/2$ electrons form a spin triplet, the
spin-correlation will be $<{\bf S}_1\cdot {\bf S}_2> =1/4$. The
rather high positive value of spin-correlation indicates that
electrons localized in QDs still have high probability to form a
spin triplet even though they are coupled with the electrons in the
leads in the Kondo regime. By increasing the interdot coupling
$t_c$, there exhibits a quantum phase transition from the triplet
state to singlet state in the ground state. The spin correlation
approach a negative value $<{\bf S}_1\cdot {\bf S}_2> \approx
-0.50$, as we know that for two electrons forming an ideal spin
singlet$<{\bf S}_1\cdot {\bf S}_2> =-0.75$, therefore the electrons
in DQDs is largely in a singlet state. In order to determine the
critical value of $t_c$, we have calculated the spin correlation
$<{\bf S}_1\cdot {\bf S}_2> $ at zero temperature for different
value of $t_c$, the result is shown in Fig.5(b). We find that, at
the quantum critical point $t_c\approx 0.7$, there is an abrupt jump
of the spin correlation $<{\bf S}_1\cdot {\bf S}_2> $. It indicates
that the quantum phase transition from triplet to singlet state is
of first order kind. According to a previous study on two-impurity
Kondo model\cite{Vojta}, we may expect that in the case of DQDs with
energy level difference, this kind of first order transition will
become Kosterlitz-Thouless type. It is easy to understand that the
exact quantum critical value of $t_c$ will depends on the the
interaction parameters, such as the on-site Coulomb repulsion $U$,
interdot capacitive coupling $V$ and the energy level $\epsilon_d$,
etc.

\section{summary}

 In summary, we have studied the ground state and the electron
transport properties of the system with DQDs in parallel
configuration. The strong on-site Coulomb repulsion and the interdot
capacitive coupling is taken into account by the nonperturbative NRG
technique. It is shown that the large interdot tunneling will
drastically change the transport properties in this system. The
ground state of DQDs exhibits a quantum phase transition from
triplet state to singlet state by increasing the interdot tunneling
amplitude. In the case of no interdot tunneling, the linear
conductance approaches to unitary limit in the doubly occupied
regime due to the underscreened Kondo effect, whereas it is greatly
suppressed when the electrons in DQDs form an singlet state with the
interdot coupling $t_c$ being larger than the critical value. For
the DQDs with strong interdot tunneling, the Fano resonance can be
observed in the linear conductance when the system are in the mixed
valence regime. One may expect that the underscreened Kondo effect
can be observed in future experiments on DQD system without direct
interdot tunneling.

\begin{acknowledgments}{ We thank J. Mravlje, R. \v{Z}itko  and M. Vojta for helpful
communication.  This project is supported by the National Natural
Science Foundation of China, the Shanghai Pujiang Program, and
Program for New Century Excellent Talents in University (NCET). }
\end{acknowledgments}


\begin{thebibliography}{ }
\bibitem{Wiel}W. G. van der Wiel, S. De Franceschi, J. M. Elzerman,
T. Fujisawa, S. Tarucha, and L. P. Kouwenhoven, Rev. Mod. Phys. {\bf
75}, 1 (2003).
\bibitem{Blick} R. H. Blick, D. Pfannkuche, R. J. Haug, K. v. Klitzing, and K. Eberl, Phys. Rev. Lett. {\bf 80},
4032 (1998); G. Schedelbeck, W. Wegscheider, M. Bichler, and G.
Abstreiter, Science {\bf 278}, 1792 (1997); T. H. Oosterkamp, T.
Fujisawa, W. G. van der Wiel, K. Ishibashi, R. V. Hijman, S.
Tarucha, and L. P. Kouwenhoven, Nature(London) {\bf 395}, 873
(1998); H. Jeong, A. M. Chang, and M. R. Melloch, Science{\bf 293},
2221 (2001);  N. J. Crag, J. M. Taylor, E. A. Lester, C. M. Marcus,
M. P. Hanson, A. C. Gossard, Science {\bf 304}, 565 (2004).
\bibitem{Chen}J. C. Chen, A. M. Chang, and M. R. Melloch, Phys. Rev.
Lett.{\bf 92}, 176801 (2004); A. W. Holleitner, R. H. Blick, A. K.
H$\ddot u$ttel, K. Eberl, J.P. Kotthaus, Science {\bf 297}, 70
(2002); A. W. Holleitner, C. R. Decker, H. Qin, K. Eberl, and R.H.
Blick, Phys. Rev. Lett. {\bf 87}, 256802 (2001).
\bibitem{Georges}A. Georges and Y. Meir, Phys. Rev. Lett. {\bf 82},
3508 (1999).
\bibitem{Aguado} R. Aguado and D. C. Langreth, Phys. Rev. Lett. {\bf
85}, 1946 (2000).
\bibitem{Lopez} R. L\'opez, R. Aguado, and G. Platero, Phys.
Rev. Lett. {\bf 89}, 136802 (2002).
\bibitem{Mravlje} J. Mravlje, A. Ram\v{s}ak, and T. Rejec, Phys. Rev. B
{\bf 73}, 241305(R) (2006).
\bibitem{Borda} L. Borda, G. Zar\'and, W. Hofstetter, B.I. Halperin,
and J.V. Delft, Phys. Rev. Lett. {\bf 90}, 026602 (2003).
\bibitem{Galpin} M. R. Galpin, D. E. Logan, and H. R. Krishnamurthy,
Phys. Rev. Lett.  {\bf 94}, 186406 (2005).
\bibitem{Martins} G. B. Martins, C. A. B\"usser, K.A. Al-Hassanieh,
A. Moreo, and E. Dagotto, Phys. Rev. Lett. {\bf 94}, 026804 (2005).
\bibitem{Guevara} M. L. L. de Guevara, F. Claro, and P. A. Orellana,
Phys.  Rev. B {\bf 67}, 195335 (2003).
\bibitem{Ding} G. H. Ding, C. K. Kim, and K. Nahm, Phys. Rev. B {\bf 71},
205313 (2005).
\bibitem{Tanaka} Y. Tanaka and N. Kawakami, Phys. Rev. B {\bf 72}, 085304
(2005).
\bibitem{Wilson} K. G. Wilson, Rev. Mod. Phys. {\bf 47}, 773 (1975).
\bibitem{Krishna-murthy} H. R. Krishna-murthy, J. W. Wilkins, and K. G. Wilson,
Phys. Rev. B{\bf 21}, 1003 (1980); Phys. Rev. B {\bf 21}, 1044
(1980).
\bibitem{Costic} T. A. Costic, A. C. Hewson, and V. Zlati\'{c}, J.
Phys. Condens. Matter {\bf 6}, 2519 (1994).
\bibitem{Bulla} R. Bulla, T. A. Costi, and T. Pruschke, Rev. Mod. Phys.
{\bf 80}, 395 (2008).
\bibitem{Izumida} W. Izumida, O. Sakai, Phys. Rev. B {\bf 62}, 10260 (2000).
\bibitem{Hofstetter} W. Hofstetter and H. Schoeller, Phys. Rev. Lett. {\bf  88},
016803 (2002); W. Hofstetter and G. Zarand, Phys. Rev. B {\bf 69},
235301 (2004).
\bibitem{Chung} C. H. Chung and W. Hofstetter, Phys. Rev. B {\bf 76},
045329 (2007).
\bibitem{Cornaglia} P. S. Cornaglia and D. R. Grempel, Phys. Rev. B {\bf 71},
075305 (2005).
\bibitem{Zitko} R. \v{Z}itko and J. Bon\v{c}a, Phys. Rev. B {\bf
73}, 035332 (2006).
\bibitem{Zitko1} R. \v{Z}itko and J. Bon\v{c}a, Phys. Rev. B {\bf
76}, 241305 (R) (2007).
\bibitem{Meir} Y. Meir, N. S. Wingreen, and P. A. Lee, Phys. Rev.
Lett. {\bf 70}, 2601 (1993).
\bibitem{Nishimoto} S. Nishimoto, T. Pruschke, and R. M. Noack, J.
Phys: Condensed Matter {\bf 18}, 981 (2006).
\bibitem{Zitko2} R. \v{Z}itko and J. Bon\v{c}a, Phys. Rev. B {\bf 74}, 045312 (2006).
\bibitem{Vojta} M. Vojta, R. Bullam and W. Hofstetter, Phys. Rev. B
{\bf 65}, 140405 (R) (2002).
\end{thebibliography}
\end{document}